\documentstyle[preprint,aps]{revtex}

\begin{document}
\author{Efrain J. Ferrer and Vivian de la Incera}
\title{Yukawa Coupling Contribution to Magnetic Field Induced Dynamical Mass }
\address{State University of New York at Fredonia, Fredonia, New York 14063\\
SUNY-FRE-98-11}
\maketitle

\begin{abstract}
By solving the gap equation in the quenched, ladder approximation for an
Abelian gauge model with Yukawa interaction in the presence of a constant
magnetic field, we show that the Yukawa interactions enhance the dynamical
generation of fermion mass. The theory is then studied at finite
temperature, where we prove that the critical magnetic field, required for
the mass generation to be important at temperatures comparable to the
electroweak critical temperature, can be substantially decreased due to the
Yukawa coupling. 
\end{abstract}

\section{Introduction}

In recent years the phenomenon of dynamical symmetry breaking in the
presence of an external magnetic field\cite{mirans} has attracted a great
deal of attention \cite{mirans}-\cite{vv1}. The essence of this effect lies
in the dimensional reduction in the dynamics of fermion pairing in the
presence of a magnetic field\cite{mirans}$.$ Due to such a dimensional
reduction, the magnetic field catalyses the dynamical generation of a
fermion condensate and a fermion mass, even in the weakest attractive
interaction between fermions.

The field-induced dynamical generation of mass (FDGM) has been found in many
models of field theories, in 2+1 and in 3+1 dimensions. The universality of
this phenomenon makes it of interest for a wide set of applications,
including problems in different areas such as cosmology\cite{lee},\cite
{shoko},\cite{vv1}, astrophysics\cite{note}, and planar condensed matter\cite
{mavro}\cite{semenof}.

In their original papers\cite{mirans}, Gusynin , Miransky and Shovkovy
suggested that the structure of the electroweak phase transition could be
affected by the FDGM. Considering FDGM in QED$_{4}$ at finite temperature,
Lee, Leung and Ng\cite{lee}$,$ and independently, Gusynin and Shovkovy\cite
{shoko}, calculated the critical temperature at which the fermion mass (and
hence the fermion condensate) evaporates. From their results one might
conclude that the dynamical generation of mass induced by a magnetic field
should play no role in the electroweak phase transition, because for it to
exist at the high temperatures typical of the electroweak scale, it would
require a primordial magnetic field too large ($\sim 10^{42}G$) to be
realistically attainable. However, as we have argued in a previous paper\cite
{vv1}, it is reasonable to expect an essential modification in the order of
the critical field when the FDGM takes place in the context of the
electroweak model, since a richer set of interactions enters in scene there.
As we will show below, this is indeed the case even in a model much simpler
than the electroweak theory. By studying a toy model similar to QED$_{4}$,
but including a Yukawa term, we prove that the Yukawa interactions can
substantially enhance the FDGM phenomenon.

Any cosmological application of the FDGM has to assume that primordial
magnetic fields could have been present during the early times of universe
evolution. Nowadays astronomical observations seems to support this
assumption. As it is known, primordial magnetic fields may be needed to
explain the large-scale galactic magnetic fields $\sim 10^{-6}G$ observed in
our own, as well as in other galaxies. The observed galactic fields have
been a source of motivation for many works on primordial-field generating
mechanisms \cite{gen-mech}. Typically, fields as large as $10^{24}G$ are
predicted by these mechanisms during the electroweak phase transition.
Moreover, Ambjorn and Olesen\cite{olesen} have claimed that seed primordial
fields even larger, $\sim 10^{33}G,$ would be necessary at the electroweak
scale to explain the observed galactic fields.

From the above discussion it can be understood that the significance of the
FDGM at the electroweak scale needs yet to be elucidated. The present paper
is a step in that direction. We study the FDGM in an Abelian gauge model of
massless fermions with a Yukawa term. This model, although simple, retains
some of the attributes of the electroweak theory. We explicitly show that
Yukawa interactions enhance the generation of mass in the presence of a
magnetic field, decreasing the critical field needed for the FDGM to be
important at temperatures comparable to the electroweak critical
temperature. For a Yukawa coupling of order of the top-quark coupling, the
critical field strength is decreased in 10 orders of magnitude as compared
to the corresponding field strength in QED.

The plan of the paper is the following. In Sec. II, we derive the
Schwinger-Dyson equation for the fermion self-energy in the quenched, ladder
approximation, solving the corresponding gap equation in the infrared
region. Thermal effects are considered in Sec. III, where we calculate the
critical temperature at which the field-induced fermion mass disappears and
estimate the order of the critical field required for the FDGM to be
significant at the electroweak scale. We discuss the implications of our
results and state our conclusions in Sec. IV. The solution of the SD
equation for the coefficient $Z_{\Vert }$ of the mass operator is derived in
the Appendix.

\section{Abelian Gauge Model with Yukawa Interaction}

Let us consider an Abelian gauge model with a Yukawa interaction described
by Lagrangian density

\begin{equation}
L=-\frac{1}{4}F^{\mu \nu }F_{\mu \nu }+i\overline{\psi }\gamma ^{\mu
}\partial _{\mu }\psi -g\overline{\psi }\gamma ^{\mu }\psi A_{\mu }-\frac{1}{%
2}\xi (\partial _{\mu }A^{\mu })^{2}+\frac{1}{2}\partial _{\mu }\phi
\partial ^{\mu }\phi -\frac{\lambda }{4}\phi ^{4}-\sqrt{2}\lambda _{y}\phi 
\overline{\psi }\psi  \label{e1}
\end{equation}

Note that $L$ has a U(1) gauge symmetry and a fermion number global
symmetry, but it does not have a continuous chiral symmetry. We consider the
present model in the presence of a constant and uniform external magnetic
field $H$. Our aim is to investigate the gap equation of the theory. We need
to go beyond a perturbative calculation to find a non zero solution of the
gap equation. In the small coupling regime of the theory a consistent, non
perturbative calculation can be carried out by using the quenched, ladder
approximation. This approximation has produced gauge invariant\cite{vv} non
trivial solutions of the gap equation in other theories with external fields%
\cite{mirans}, so it is natural to expect that it will yield similar results
in the present case.

The gap equation of the theory described by the Lagrangian density $\left( 
\text{\ref{e1}}\right) $ can be obtained from the Schwinger-Dyson equation
for the fermion self-energy 
\[
\overline{G}^{-1}(x,y)=G^{-1}(x,y)+ig\int d^{4}ud^{4}w\gamma \overline{G}%
(x,u)\overline{D}(x-w)\overline{\Gamma }_{\overline{\psi }\psi A}(u,y,w)
\]
\begin{equation}
+i\sqrt{2}\lambda _{y}\int d^{4}ud^{4}w\overline{G}(x,u)\overline{S}(x-w)%
\overline{\Gamma }_{\overline{\psi }\psi \phi }(u,y,w)  \label{e2}
\end{equation}

Here $G$ refers to fermion propagators, and $D$ and $S$ to gauge and scalar
boson propagators respectively. $\overline{\Gamma }_{\overline{\psi }\psi A}$
and $\overline{\Gamma }_{\overline{\psi }\psi \phi }$ are three-fields
vertex functions. The bar indicates full Green functions. A compact notation
where tensorial and spinorial indexes have been suppressed is understood.

After taking the quenched, ladder approximation (where the fermion
propagator in the presence of an external magnetic field is taken full,
while the vertices, as well as the gauge and scalar boson propagators, are
taken bare) of Eq. (\ref{e2}), one obtains the following equation for the
fermion mass operator

\begin{eqnarray}
M(x,y) &=&\overline{G}^{-1}(x,y)-G^{-1}(x,y)=-ig^{2}\int d^{4}ud^{4}w\gamma 
\overline{G}(x,u)\gamma D(x-w)  \nonumber \\
&&-i\left( \sqrt{2}\lambda _{y}\right) ^{2}\int d^{4}ud^{4}w\overline{G}%
(x,u)S(x-w)  \label{e2-1}
\end{eqnarray}

To solve Eq. $\left( \text{\ref{e2-1}}\right) $ we need to transform it to
momentum space. However, it is known that the Fourier transform of the
fermion Green function in the presence of a magnetic field does not yield a
diagonal-in-p function. The reason is that in the problem with external
field the fermion asymptotic states are no longer plane waves. A suitable
solution to this technical problem was found many years ago by Schwinger\cite
{schwinger}, who introduced the proper-time representation of the fermion
Green function in the presence of a constant external field. Here, however,
we prefer to adopt another approach due to Ritus\cite{ritus}. Ritus' method
is based on the use of a representation (known as the $E_{p}$
representation) spanned by the solutions $\psi _{p}$ of the eigenvalue
equation ($\gamma \Pi )^{2}\psi _{p}=\overline{p}^{2}\psi _{p}$. The
operator ($\gamma \Pi )^{2}=(\gamma ^{\mu }(i\partial _{\mu }-gA_{\mu }))^{2}
$commutes with the mass operator. Hence, the $\psi _{p}$ can be used to
generate a set of complete and orthonormal eigenfunction-matrices of the
mass operator. In the chiral representation of the Dirac matrices, where $%
\gamma _{5}$ and $%
\mathop{\textstyle \sum }%
_{3}=i\gamma _{1}\gamma _{2}$ are both diagonal, the $\psi _{p}$ take the
form

\begin{equation}
\psi _{p}=E_{p\sigma \chi }(x)\omega _{\sigma \chi }  \label{n1}
\end{equation}
The bispinors $\omega _{\sigma \chi }$ are eigenvectors of $\gamma _{5}$ and 
$%
\mathop{\textstyle \sum }%
_{3}$, with eigenvalues $\chi =\pm 1$ and $\sigma =\pm 1,$ respectively.

In the case of a purely magnetic field background (crossed field case)
directed along the z-direction, the $E_{p\sigma \chi }$ functions are

\begin{equation}
E_{p\sigma }(x)=N(n)e^{i(p_{0}x^{0}+p_{2}x^{2}+p_{3}x^{3})}D_{n}(\rho )
\label{n2}
\end{equation}
where $D_{n}(\rho )$ are the parabolic cylinder functions\cite{handbook}
with argument $\rho =\sqrt{2\left| gH\right| }(x_{1}-\frac{p_{2}}{gH})$ and
positive integer index

\begin{equation}
n=n(k,\sigma )\equiv k+\frac{gH\sigma }{2\left| gH\right| }-\frac{1}{2}%
,\quad n=0,1,2,...  \label{n3}
\end{equation}
$N(n)=(4\pi \left| gH\right| )^{\frac{1}{4}}/\sqrt{n!}$ is a normalization
factor. Here $p$ represents the set $(p_{0},p_{2,}p_{3},k),$ which
determines the eigenvalue $\overline{p}^{2}=-p_{0}^{2}+p_{3}^{2}+2\left|
gH\right| k$ in ($\gamma \Pi )^{2}\psi _{p}=\overline{p}^{2}\psi _{p}.$ Note
that in this case $E_{p\sigma \chi }$ does not depend on $\chi $.

The $E_{p}$ representation is obtained forming the eigenfunction-matrices

\begin{equation}
E_{p}(x)=\sum\limits_{\sigma }E_{p\sigma }(x)\Delta (\sigma ),  \label{n4}
\end{equation}
where 
\begin{equation}
\Delta (\sigma )=diag(\delta _{\sigma 1},\delta _{\sigma -1},\delta _{\sigma
1},\delta _{\sigma -1}),\qquad \sigma =\pm 1,  \label{n5}
\end{equation}

It is easy to check that the $E_{p}$ functions are orthonormal

\begin{equation}
\int d^{4}x\overline{E}_{p^{\prime }}(x)E_{p}(x)=(2\pi )^{4}\widehat{\delta }%
^{(4)}(p-p^{\prime })\equiv (2\pi )^{4}\delta _{kk^{\prime }}\delta
(p_{0}-p_{0}^{\prime })\delta (p_{2}-p_{2}^{\prime })\delta
(p_{3}-p_{3}^{\prime })  \label{n6}
\end{equation}
as well as complete

\begin{equation}
\sum\limits_{k}\int \ dp_{0}dp_{2}dp_{3}E_{p}(x)\overline{E}_{p}(y)=(2\pi
)^{4}\delta ^{(4)}(x-y)  \label{n7}
\end{equation}
Here we have used $\overline{E}_{p}(x)=\gamma ^{0}E_{p}^{\dagger }\gamma
^{0}.$

They also satisfy two important relations

\begin{equation}
\gamma \cdot \Pi E_{p}(x)=E_{p}(x)\gamma \cdot \overline{p}  \label{n8}
\end{equation}

\begin{equation}
\int d^{4}x^{\prime }M(x,x^{\prime })E_{p}(x)=E_{p}(x)\widetilde{%
\mathop{\textstyle \sum }%
}_{A}(\overline{p})  \label{n9}
\end{equation}
where $\widetilde{%
\mathop{\textstyle \sum }%
}_{A}(\overline{p})$ is the fermion mass operator in momentum coordinates.

In the $E_{p}$ representation the SD equation $\left( \text{\ref{e2-1}}%
\right) $ for the mass operator becomes

\[
(2\pi )^{4}\delta _{kk^{\prime }}\delta (p_{0}-p_{0}^{\prime })\delta
(p_{2}-p_{2}^{\prime })\delta (p_{3}-p_{3}^{\prime })\widetilde{%
\mathop{\textstyle \sum }%
}_{A}(\overline{p}) 
\]

\[
=-ig^{2}\int d^{4}xd^{4}x^{\prime }\sum_{k"}\int \frac{dp"_{0}dp"_{2}dp"_{3}%
}{(2\pi )^{4}}\{\overline{E}_{p}(x)\gamma ^{\mu }E_{p"}(x)\frac{1}{\gamma
\cdot \overline{p}"-\widetilde{%
\mathop{\textstyle \sum }%
}_{A}(\overline{p}")} 
\]
\[
\times \overline{E}_{p"}(x^{\prime })\gamma ^{\nu }E_{p^{\prime }}(x^{\prime
})D_{\mu \nu }(x-x^{\prime })\}-i2\lambda _{y}^{2}\int d^{4}xd^{4}x^{\prime
}\sum_{k"}\int \frac{dp"_{0}dp"_{2}dp"_{3}}{(2\pi )^{4}}\{\overline{E}%
_{p}(x)E_{p"}(x) 
\]
\begin{equation}
\times \frac{1}{\gamma \cdot \overline{p}"-\widetilde{%
\mathop{\textstyle \sum }%
}_{A}(\overline{p}")}\overline{E}_{p"}(x^{\prime })E_{p^{\prime }}(x^{\prime
})S(x-x^{\prime })\}  \label{e3}
\end{equation}
with 
\begin{equation}
D_{\mu \nu }(x-x^{\prime })=-\int \frac{d^{4}q}{(2\pi )^{4}}\frac{%
e^{-iq\cdot (x-x^{\prime })}}{q^{2}-i\epsilon }\left( g_{\mu \nu }-(1-\xi )%
\frac{q_{\mu }q_{\nu }}{q^{2}-i\epsilon }\right)  \label{n10}
\end{equation}
the bare photon propagator, and

\begin{equation}
S(x-x^{\prime })=\int \frac{d^{4}q}{(2\pi )^{4}}\frac{e^{-iq\cdot
(x-x^{\prime })}}{q^{2}-i\epsilon }  \label{n11}
\end{equation}
the bare scalar propagator.

Using the properties of the parabolic cylinder functions and Eqs. $\left( 
\text{\ref{n8}}\right) $-$\left( \text{\ref{n9}}\right) $, the integrals in
x and x' in Eq. (\ref{e3}), as well as the integrals in $p_{0,\text{ }}p_{2},
$and $p_{3\text{,}}$ can be done yielding 
\[
\delta _{kk^{\prime }}\widetilde{%
\mathop{\textstyle \sum }%
}_{A}(\overline{p})=ig^{2}2\left| gH\right| \sum_{k"}\sum_{\{\sigma \}}\int 
\frac{d^{4}\widehat{q}}{(2\pi )^{4}}\{\frac{e^{isgn(gH)(n-n"+\widetilde{n}%
"-n^{\prime })\varphi }}{\sqrt{n!n"!\widetilde{n}"!n^{\prime }!}}e^{-%
\widehat{q}_{\bot }^{2}}J_{nn"}(\widehat{q}_{\bot })J_{\widetilde{n}%
"n^{\prime }}(\widehat{q}_{\bot })\frac{1}{\widehat{q}^{2}}
\]
\[
\times \left( g_{\mu \nu }-\left( 1-\xi \right) \frac{\widehat{q}_{\mu }%
\widehat{q}_{\nu }}{\widehat{q}^{2}}\right) \Delta \gamma ^{\mu }\Delta "%
\frac{1}{\gamma \cdot \overline{p}"-\widetilde{%
\mathop{\textstyle \sum }%
}_{A}(\overline{p}")}\tilde{\Delta}"\gamma ^{\nu }\Delta ^{\prime }\}
\]
\[
-i2\lambda _{y}^{2}(2\left| gH\right| )\sum_{k"}\sum_{\{\sigma \}}\int \frac{%
d^{4}\widehat{q}}{(2\pi )^{4}}\{\frac{e^{isgn(gH)(n-n"+\widetilde{n}%
"-n^{\prime })\varphi }}{\sqrt{n!n"!\widetilde{n}"!n^{\prime }!}}e^{-%
\widehat{q}_{\bot }^{2}}J_{nn"}(\widehat{q}_{\bot })J_{\widetilde{n}%
"n^{\prime }}(\widehat{q}_{\bot })\frac{1}{\widehat{q}^{2}}
\]
\begin{equation}
\times \Delta \Delta "\frac{1}{\gamma \cdot \overline{p}"-\widetilde{%
\mathop{\textstyle \sum }%
}_{A}(\overline{p}")}\tilde{\Delta}"\Delta ^{\prime }\}  \label{e6}
\end{equation}
where the following notation has been used

\begin{equation}
J_{n_{p}n_{r}}(\widehat{q}_{\perp })\equiv \sum\limits_{m=0}^{\min
(n_{p},n_{r})}\frac{n_{p}!n_{r}!}{m!(n_{p}-m)!(n_{r}-m)!}[isgn(gH)\widehat{q}%
_{\perp }]^{n_{p}+n_{r}-2m}  \label{e6-a}
\end{equation}
\begin{equation}
\overline{p}"\equiv (p_{0}-q_{0},0,-sgn(gH)\sqrt{2\left| gH\right| k"}%
,p_{3}-q_{3})  \label{e6-d}
\end{equation}
\begin{equation}
\sum_{\{\sigma \}}\equiv \sum_{\sigma \sigma "\sigma ^{\prime }\widetilde{%
\sigma }"}  \label{ex1}
\end{equation}
\[
n\equiv n(k,\sigma ),\qquad n^{\prime }\equiv n(k^{\prime },\sigma ^{\prime
}),\qquad n"\equiv n(k",\sigma "),\qquad \widetilde{n}"\equiv n(k",%
\widetilde{\sigma }"), 
\]
and the dimensionless variables

\begin{equation}
\widehat{q}_{\mu }\equiv \frac{q_{\mu }\sqrt{2\left| gH\right| }}{2gH}\
,\qquad \mu =0,1,2,3  \label{e6-b}
\end{equation}
and polar coordinates for the transverse components of $\widehat{q}_{\mu }$

\begin{equation}
\widehat{q}_{\perp }\equiv \sqrt{\widehat{q}_{1}^{2}+\widehat{q}_{2}^{2}}%
,\quad \varphi \equiv \arctan (\widehat{q}_{2}/\widehat{q}_{1})  \label{e6-c}
\end{equation}
have been introduced.

The first term of the right hand side (RHS) in Eq. (\ref{e6}) coincides with
the result previously found in references [3] for the pure QED case. Because
of the presence of the Yukawa interaction in the present model, we have
obtained an additional contribution (the second integral in the RHS of Eq. (%
\ref{e6})) whose consequences for the gap equation will not be trivial, as
we show below.

To solve equation (\ref{e6}), we need the mass operator structure. In the
presence of the external magnetic field the mass operator structure is quite
rich\cite{vv}$.$ However, as argued in a previous paper\cite{vv}$,$ within
the present approximation a simpler structure can be used. Thus, we consider

\begin{equation}
\widetilde{%
\mathop{\textstyle \sum }%
}_{A}(\overline{p})=Z_{_{\Vert }}(\overline{p})\gamma \cdot \overline{p}%
_{_{\Vert }}+Z_{\bot }(\overline{p})\gamma \cdot \overline{p}_{_{\bot }}+m(%
\overline{p})  \label{e8}
\end{equation}
Note the separation between transverse and perpendicular variables.

After substituting the above structure for $\widetilde{%
\mathop{\textstyle \sum }%
}_{A}(\overline{p})$, Eq. (\ref{e6}) can be written as

\[
\delta _{kk^{\prime }}\left[ Z_{_{\Vert }}\gamma \cdot \overline{p}_{_{\Vert
}}+Z_{\bot }\gamma \cdot \overline{p}_{_{\bot }}+m(\overline{p})\right] 
\]
\[
=ig^{2}2\left| gH\right| \sum_{k"}\sum_{\{\sigma \}}\int \frac{d^{4}\widehat{%
q}}{(2\pi )^{4}}\frac{e^{isgn(gH)(n-n"+\widetilde{n}"-n^{\prime })\varphi }}{%
\sqrt{n!n"!\widetilde{n}"!n^{\prime }!}}e^{-\widehat{q}_{\bot }^{2}}J_{nn"}(%
\widehat{q}_{\bot })J_{\widetilde{n}"n^{\prime }}(\widehat{q}_{\bot })\frac{1%
}{\widehat{q}^{2}} 
\]
\[
\times \left\{ \left( g_{\mu \nu }-\left( 1-\xi \right) \frac{\widehat{q}%
_{\mu }\widehat{q}_{\nu }}{\widehat{q}^{2}}\right) \Delta \gamma ^{\mu
}\Delta "\frac{m(\overline{p}")-(1+Z_{_{\Vert }})\gamma \cdot \overline{p}%
_{_{\Vert }}"-(1+Z_{\bot })\gamma \cdot \overline{p}_{_{\bot }}"}{%
(1+Z_{_{\Vert }})^{2}\overline{p}"^{2}+(1+Z_{\bot })^{2}\overline{p}_{_{\bot
}}^{2}+m^{2}(\overline{p}")}\tilde{\Delta}"\gamma ^{\nu }\Delta ^{\prime
}\right\} 
\]
\[
-i2\lambda _{y}^{2}(2\left| gH\right| )\sum_{k"}\sum_{\{\sigma \}}\int \frac{%
d^{4}\widehat{q}}{(2\pi )^{4}}\frac{e^{isgn(gH)(n-n"+\widetilde{n}%
"-n^{\prime })\varphi }}{\sqrt{n!n"!\widetilde{n}"!n^{\prime }!}}e^{-%
\widehat{q}_{\bot }^{2}}J_{nn"}(\widehat{q}_{\bot })J_{\widetilde{n}%
"n^{\prime }}(\widehat{q}_{\bot })\frac{1}{\widehat{q}^{2}} 
\]

\begin{equation}
\times \left\{ \Delta \Delta "\frac{m(\overline{p}")-(1+Z_{_{\Vert }})\gamma
\cdot \overline{p}_{_{\Vert }}"-(1+Z_{\bot })\gamma \cdot \overline{p}%
_{_{\bot }}"}{(1+Z_{_{\Vert }})^{2}\overline{p}"^{2}+(1+Z_{\bot })^{2}%
\overline{p}_{_{\bot }}^{2}+m^{2}(\overline{p}")}\tilde{\Delta}"\Delta
^{\prime }\right\}  \label{n12}
\end{equation}
It can be further simplified taking into account that large $\widehat{q}%
_{\bot }$ contributions to Eq. (\ref{n12}) are suppressed by the factor $e^{-%
\widehat{q}_{\bot }^{2}}$, and therefore we can approximate the $%
J_{nn^{\prime }}(\widehat{q}_{\bot })$ functions by their infrared behavior

\begin{equation}
J_{nn^{\prime }}(\widehat{q}_{\bot })\approx n!\delta _{nn^{\prime }}
\label{n13}
\end{equation}

This approximation allows us to eliminate the $\varphi $ dependence in the
integrand to obtain

\[
\delta _{kk^{\prime }}\left[ Z_{_{\Vert }}\gamma \cdot \overline{p}_{_{\Vert
}}+Z_{\bot }\gamma \cdot \overline{p}_{_{\bot }}+m(\overline{p})\right] 
\]
\[
=ig^{2}2\left| gH\right| \sum_{k"}\sum_{\{\sigma \}}\delta _{nn"}\delta _{%
\widetilde{n}"n^{\prime }}\int \frac{d^{4}\widehat{q}}{(2\pi )^{4}}\frac{e^{-%
\widehat{q}_{\bot }^{2}}}{\widehat{q}^{2}}\left( \frac{1}{(1+Z_{_{\Vert
}})^{2}\overline{p}"^{2}+(1+Z_{\bot })^{2}\overline{p}_{_{\bot }}^{2}+m^{2}(%
\overline{p}")}\right) 
\]
\begin{eqnarray*}
&&\times \{m(\overline{p}")\left[ \Delta \gamma ^{\mu }\Delta "\tilde{\Delta}%
"\gamma ^{\nu }\Delta ^{\prime }-\frac{\left( 1-\xi \right) }{\widehat{q}^{2}%
}\Delta (\gamma \cdot \widehat{q})\Delta "\tilde{\Delta}"(\gamma \cdot 
\widehat{q})\Delta ^{\prime }\right] \\
&&-(1+Z_{_{\Vert }})\left[ \Delta \gamma ^{\mu }\Delta "\left( \gamma \cdot 
\overline{p}_{_{\Vert }}"\right) \tilde{\Delta}"\gamma _{\mu }\Delta
^{\prime }-\frac{\left( 1-\xi \right) }{\widehat{q}^{2}}\Delta (\gamma \cdot 
\widehat{q})\Delta "\left( \gamma \cdot \overline{p}_{_{\Vert }}"\right) 
\tilde{\Delta}"(\gamma \cdot \widehat{q})\Delta ^{\prime }\right] \\
&&-(1+Z_{\bot })\left[ \Delta \gamma ^{\mu }\Delta "\left( \gamma \cdot 
\overline{p}_{_{\bot }}"\right) \tilde{\Delta}"\gamma _{\mu }\Delta ^{\prime
}-\frac{\left( 1-\xi \right) }{\widehat{q}^{2}}\Delta (\gamma \cdot \widehat{%
q})\Delta "\left( \gamma \cdot \overline{p}_{_{\bot }}"\right) \tilde{\Delta}%
"(\gamma \cdot \widehat{q})\Delta ^{\prime }\right] \}
\end{eqnarray*}
\begin{eqnarray}
&&-i2\lambda _{y}^{2}(2\left| gH\right| )\sum_{k"}\sum_{\{\sigma \}}\delta
_{nn"}\delta _{\widetilde{n}"n^{\prime }}\int \frac{d^{4}\widehat{q}}{(2\pi
)^{4}}\frac{e^{-\widehat{q}_{\bot }^{2}}}{\widehat{q}^{2}}  \nonumber \\
&&\times \left( \frac{m(\overline{p}")\Delta \Delta "\tilde{\Delta}"\Delta
^{\prime }-(1+Z_{_{\Vert }})\Delta \Delta "\left( \gamma \cdot \overline{p}%
_{_{\Vert }}"\right) \tilde{\Delta}"\Delta ^{\prime }-(1+Z_{\bot })\Delta
\Delta "\left( \gamma \cdot \overline{p}_{_{\bot }}"\right) \tilde{\Delta}%
"\Delta ^{\prime }}{(1+Z_{_{\Vert }})^{2}\overline{p}"^{2}+(1+Z_{\bot })^{2}%
\overline{p}_{_{\bot }}^{2}+m^{2}(\overline{p}")}\right)  \label{n14}
\end{eqnarray}

To perform the summation in the spin indices we can make use of the
following relations satisfied by the $\Delta $ matrices $\left( \text{\ref
{n5}}\right) $%
\begin{equation}
\Delta (\sigma )\Delta (\sigma ^{\prime })=\delta _{\sigma \sigma ^{\prime
}}\Delta (\sigma )  \label{n15}
\end{equation}
\[
\Delta (1)+\Delta (-1)=I 
\]

\begin{equation}
\Delta \gamma _{\parallel }^{\mu }=\gamma _{\parallel }^{\mu }\Delta ,\qquad
\qquad \qquad with\ \text{\quad }\gamma _{\parallel }^{\mu }=(\gamma
^{0},0,0,\gamma ^{3})  \label{n16}
\end{equation}

\begin{center}
\begin{equation}
\Delta \gamma _{\perp }^{\mu }=\gamma _{\perp }^{\mu }(1-\Delta ),\qquad
\qquad with\ \text{\quad }\gamma _{\perp }^{\mu }=(0,\gamma ^{1},\gamma
^{2},0)  \label{n17}
\end{equation}
\end{center}

to find that

\begin{equation}
\sum_{\{\sigma \}}\delta _{nn"}\delta _{\widetilde{n}"n^{\prime }}\Delta
\Delta "\tilde{\Delta}"\Delta ^{\prime }=\delta _{kk^{\prime }}\delta _{kk"}
\label{n18}
\end{equation}
\begin{equation}
\sum_{\{\sigma \}}\delta _{nn"}\delta _{\widetilde{n}"n^{\prime }}\Delta
\gamma ^{\mu }\Delta "\left( \gamma \cdot \overline{p}_{\perp }"\right) 
\tilde{\Delta}"\gamma _{\mu }\Delta ^{\prime }=2\left( \gamma \cdot 
\overline{p}_{\perp }"\right) \delta _{kk^{\prime }}\delta _{kk"}
\label{n19}
\end{equation}
\begin{equation}
\sum_{\{\sigma \}}\delta _{nn"}\delta _{\widetilde{n}"n^{\prime }}\Delta
\gamma ^{\mu }\Delta "\left( \gamma \cdot \overline{p}_{_{\Vert }}"\right) 
\tilde{\Delta}"\gamma _{\mu }\Delta ^{\prime }=2\left( \gamma \cdot 
\overline{p}_{_{\Vert }}"\right) \delta _{kk^{\prime }}\left( \delta
_{k,k"-sgn(gH)}\Delta (1)+\delta _{k,k"+sgn(gH)}\Delta (-1)\right)
\label{n20}
\end{equation}
\begin{equation}
\sum_{\{\sigma \}}\delta _{nn"}\delta _{\widetilde{n}"n^{\prime }}\Delta
\left( \gamma \cdot \widehat{q}\right) \Delta "\left( \gamma \cdot \overline{%
p}_{\perp }"\right) \tilde{\Delta}"\left( \gamma \cdot \widehat{q}\right)
\Delta ^{\prime }\simeq \left( \gamma \cdot \overline{p}_{\perp }"\right)
\delta _{kk^{\prime }}\delta _{kk"}  \label{n21}
\end{equation}
\begin{equation}
\sum_{\{\sigma \}}\delta _{nn"}\delta _{\widetilde{n}"n^{\prime }}\frac{%
\Delta \left( \gamma \cdot \widehat{q}\right) \Delta "\left( \gamma \cdot 
\overline{p}_{_{\Vert }}"\right) \tilde{\Delta}"\left( \gamma \cdot \widehat{%
q}\right) \Delta ^{\prime }}{\widehat{q}^{2}}\simeq \delta _{kk^{\prime
}}\delta _{kk"}\frac{\left( \gamma \cdot \widehat{q}_{_{\Vert }}\right)
\left( \gamma \cdot \overline{p}_{_{\Vert }}"\right) \left( \gamma \cdot 
\widehat{q}_{_{\Vert }}\right) }{\widehat{q}^{2}}  \label{n22}
\end{equation}
\begin{equation}
\sum_{\{\sigma \}}\delta _{nn"}\delta _{\widetilde{n}"n^{\prime }}\Delta
\Delta "\left( \gamma \cdot \overline{p}_{_{\bot }}"\right) \tilde{\Delta}%
"\Delta ^{\prime }=\delta _{kk^{\prime }}\delta _{kk"}\left( \gamma \cdot 
\overline{p}_{_{\bot }}"\right)  \label{n23}
\end{equation}
\begin{equation}
\sum_{\{\sigma \}}\delta _{nn"}\delta _{\widetilde{n}"n^{\prime }}\Delta
\Delta "\left( \gamma \cdot \overline{p}_{_{\Vert }}"\right) \tilde{\Delta}%
"\Delta ^{\prime }=\delta _{kk^{\prime }}\delta _{kk"}\left( \gamma \cdot 
\overline{p}_{_{\Vert }}"\right)  \label{n24}
\end{equation}

In the above equations we dropped terms proportional to $\widehat{q}%
_{_{\perp }}$ taking into account the small $\widehat{q}_{_{\perp }}$%
approximation here considered. The remaining $\sigma $ sums were obtained in
ref. \cite{lee}. Note that the $\delta _{kk^{\prime }}$ appearing in all
terms cancels out with the one of the left hand side (LHS) of Eq. (\ref{n14}%
). The SD equation is then expressed as 
\[
Z_{_{\Vert }}\gamma \cdot \overline{p}_{_{\Vert }}+Z_{\bot }\gamma \cdot 
\overline{p}_{_{\bot }}+m(\overline{p})=i2\left| gH\right| \sum_{k"}\int 
\frac{d^{4}\widehat{q}}{(2\pi )^{4}}\frac{e^{-\widehat{q}_{\bot }^{2}}}{%
\widehat{q}^{2}}\left( \frac{1}{(1+Z_{_{\Vert }})^{2}\overline{p}%
"^{2}+(1+Z_{\bot })^{2}\overline{p}_{_{\bot }}^{2}+m^{2}(\overline{p}")}%
\right) 
\]
\begin{eqnarray}
&&\times \{g^{2}\{m(\overline{p}")\left[ \left( 1-\xi \right) \delta
_{kk"}-2\left( \delta _{kk"}+\delta _{k,k"-sgn(gH)}\Delta (1)+\delta
_{k,k"+sgn(gH)}\Delta (-1)\right) \right]   \nonumber \\
&&+(1+Z_{_{\Vert }})\left[ \left( 1-\xi \right) \delta _{kk"}\frac{\left(
\gamma \cdot \widehat{q}_{_{\Vert }}\right) \left( \gamma \cdot \overline{p}%
_{_{\Vert }}"\right) \left( \gamma \cdot \widehat{q}_{_{\Vert }}\right) }{%
\widehat{q}^{2}}-2\left( \gamma \cdot \overline{p}_{_{\Vert }}"\right)
\left( \delta _{k,k"-sgn(gH)}\Delta (1)+\delta _{k,k"+sgn(gH)}\Delta
(-1)\right) \right]   \nonumber \\
&&+(1+Z_{\bot })\left[ \left( 1-\xi \right) \delta _{kk"}\left( \gamma \cdot 
\overline{p}_{\bot }"\right) -2\delta _{kk"}\left( \gamma \cdot \overline{p}%
_{_{\bot }}"\right) \right] \}  \nonumber \\
&&-2\lambda _{y}^{2}\{m(\overline{p}")\delta _{kk"}-(1+Z_{_{\Vert }})\delta
_{kk"}\left( \gamma \cdot \overline{p}_{_{\Vert }}"\right) -(1+Z_{\bot
})\delta _{kk"}\left( \gamma \cdot \overline{p}_{\bot }\right) \}\}
\label{n25}
\end{eqnarray}

The $\Delta ^{\prime }s$ terms, appearing in the RHS of Eq. (\ref{n25}), but
not in the LHS, remind us that the mass operator structure used here, Eq. (%
\ref{e8}), is not the most general one in the presence of an external
magnetic field \cite{vv}. However, to study the dynamical generation of mass
we can restrict our calculations to certain momentum region on which the
simplified structure (\ref{n8}) gives rise to self-consistent results. Let
us recall that in the case of QED it has been shown \cite{mirans},\cite{lee},%
\cite{hong} that the dynamical generation of mass in the presence of a
magnetic field is governed by the fermion infrared dynamics. It is natural
to expect that a similar situation takes place in the model we are
considering here. Thereby, we shall restrict the above SD equation to the
lower Landau level (LLL) approximation, on which $\overline{p}_{_{\bot }}=k=0
$, which is better justified if we simultaneously assume that the external
momentum lies in the infrared region $\overline{p}^{2}<<\left| eH\right| .$
If, in addition, we use the Feynman gauge, and realize that the main
contribution to the sum in $k"$ comes from the $k"=k=0$ terms, we can
rewrite the SD equation as 
\[
Z_{_{\Vert }}\gamma \cdot \overline{p}_{_{\Vert }}+m(\overline{p}_{\Vert
})\simeq i2\left| gH\right| \int \frac{d^{4}\widehat{q}}{(2\pi )^{4}}\frac{%
e^{-\widehat{q}_{\bot }^{2}}}{\widehat{q}^{2}}\left( \frac{1}{(1+Z_{_{\Vert
}})^{2}\overline{p}"^{2}+m^{2}(\overline{p}")}\right) 
\]
\begin{equation}
\times \left\{ -2(g^{2}+\lambda _{y}^{2})m(\overline{p}")+2\lambda
_{y}^{2}(1+Z_{_{\Vert }})\delta _{kk"}\left( \gamma \cdot \overline{p}%
_{_{\Vert }}"\right) \right\}   \label{n26}
\end{equation}

Performing a Wick rotation to Euclidean space, Eq.(\ref{n26}) leads to the
following two equations, one for each independent structure,

\begin{center}
\begin{equation}
Z_{_{\Vert }}\gamma \cdot \overline{p}_{_{\Vert }}=-4\lambda _{y}^{2}(\left|
gH\right| )\int \frac{d^{4}\widehat{q}}{(2\pi )^{4}}\frac{e^{-\widehat{q}%
_{\bot }^{2}}}{\widehat{q}^{2}}\frac{(1+Z_{_{\Vert }})\gamma \cdot (%
\overline{p}_{_{\Vert }}-q_{_{_{\Vert }}})}{\left[ (1+Z_{_{\Vert }})^{2}(%
\overline{p}_{_{\Vert }}-q_{_{_{\Vert }}})^{2}+m^{2}(\overline{p}_{_{\Vert
}}-q_{_{_{\Vert }}})\right] }  \label{e9}
\end{equation}
\begin{equation}
m(\overline{p}_{_{\Vert }})=4\left| gH\right| (g^{2}+\lambda _{y}^{2})\int 
\frac{d^{4}\widehat{q}}{(2\pi )^{4}}\frac{e^{-\widehat{q}_{\bot }^{2}}m(%
\overline{p}_{_{\Vert }}-q_{_{_{\Vert }}})}{\widehat{q}^{2}\left[
(1+Z_{_{\Vert }})^{2}(\overline{p}_{_{\Vert }}-q_{_{_{\Vert }}})^{2}+m^{2}(%
\overline{p}_{_{\Vert }}-q_{_{_{\Vert }}})\right] }  \label{e10}
\end{equation}
\end{center}

As shown in the Appendix, the first equation has solution $Z_{_{\Vert }}=0.$
The second is the gap equation. Note that it is very similar to the gap
equation found by Lee, Leung and Ng \cite{lee} (after using $Z_{_{\Vert }}=0$%
), except that the coupling factor here is $(g^{2}+\lambda _{y}^{2})$
instead of just $g^{2}$. The main contributions to Eq.(\ref{e10}) come from
the infrared region $\widehat{q}_{\bot }^{2}<<\left| gH\right| ,\quad
q_{_{_{\Vert }}}^{2}<<\left| gH\right| .$ A solution to Eq. (\ref{e10}) can
be explicitly found in the infrared approximation $\overline{p}_{_{\Vert
}}\approx 0,$ taking into account that the mass parameter is dominated by
the small momenta contributions\cite{hongkim}. Then, we can approximate the
mass by its infrared value $m(\overline{p}_{_{\Vert }}-q_{_{_{\Vert
}}})\simeq m(\overline{p}_{_{\Vert }})\simeq m(0),$ and use that $Z_{_{\Vert
}}=0$ to arrive at the infrared gap equation

\begin{equation}
1\simeq \frac{\left| gH\right| (g^{2}+\lambda _{y}^{2})}{4\pi ^{2}}%
\int\limits_{0}^{\infty }d^{2}\widehat{q}_{\bot }\int\limits_{0}^{\infty
}d^{2}\widehat{q}_{_{_{\Vert }}}\frac{e^{-\widehat{q}_{\bot }^{2}}}{\widehat{%
q}_{\bot }^{2}+\widehat{q}_{_{\Vert }}^{2}}\frac{e^{-\widehat{q}_{\bot }^{2}}%
}{2\left| gH\right| \widehat{q}_{_{_{\Vert }}}{}^{2}+m^{2}},  \label{e10-1}
\end{equation}
which can be integrated to obtain the non trivial solution 
\begin{equation}
m\approx \sqrt{\left| gH\right| }\exp \left[ -\sqrt{\frac{\pi }{\frac{g^{2}}{%
4\pi }+\frac{\lambda _{y}^{2}}{4\pi }}}\right]  \label{n28}
\end{equation}

The consistency of the approximation requires $m<<\sqrt{\left| gH\right| },$
which is satisfied if $\frac{g^{2}}{4\pi }+\frac{\lambda _{y}^{2}}{4\pi }<<1,
$ so the dynamical mass appears in the weak coupling region of the theory.
Note that, because of the exponential function in Eq. $\left( \text{\ref{n28}%
}\right) $, small changes in the exponent can yield substantial changes in
the mass. For instance, for $\lambda _{y}\simeq 0.7$, a value comparable to
the top-quark Yukawa coupling, the dynamical mass (\ref{n28}) is five orders
of magnitude larger than the mass found in QED \cite{mirans}$-$\cite{lee},
which is given by $m\simeq \sqrt{\left| gH\right| }\exp \left[ -\sqrt{\frac{%
4\pi ^{2}}{g^{2}}}\right] .$

We should underline that in this model the generation of a dynamical mass
cannot be linked to the breaking of a continuous chiral symmetry, because
there is no chiral symmetry to begin with. Since our goal is to use the
present toy model to get insight of the FDGM phenomenon in the electroweak
theory, where there is no chiral symmetry to break, the present model is a
good candidate for our purpose. We must point out that, even though the
appearance of the dynamical mass in the present case can be easily traced to
the existence of a fermion-antifermion condensate\cite{note2}, there is no
Goldstone field produced in this theory, because there is no continuous
symmetry broken by the fermion condensate. We expect that this should not be
the case in the electroweak model. If a fermion condensate is catalyzed by
the magnetic field there, it might give rise to a field-dependent vev of the
scalar field and hence to a Higgs-like spontaneous gauge symmetry breaking.

Eq. $\left( \text{\ref{n28}}\right) $ clearly indicates that Yukawa
interactions enhance the dynamical generation of the fermion mass in the
presence of a magnetic field. To grasp the possible significance of this
result in the context of the electroweak phase transition, finite
temperature effects has to be considered.

\section{Finite Temperature Calculations}

Finite temperature effects can be incorporated using the well known
imaginary-time Matsubara formalism, in which the Euclidean time variable is
compactified to a circle of radius $\beta =\frac{1}{T},$ (T is the absolute
temperature), the fourth components of the momenta are consequently
discretized according to

\[
q_{4}=2n\pi /\beta \hspace{0.15in}\text{ with\qquad }n=0,\pm 1,\pm 2,...%
\text{ \qquad for bosons} 
\]
\[
q_{4}=(2n+1)\pi /\beta \hspace{0.15in}\text{ with \qquad }n=0,\pm 1,\pm 2,...%
\text{ \qquad for fermions} 
\]
and, in the functional integrals, boson (fermion) fields are periodic
(antiperiodic) in time with period $\beta .$

The calculation leading to the gap equation in the above section can be
performed at finite temperature in a similar way, taking into account that
now the integrals in the fourth components of the momenta must be
substituted by sums due to the discrete character of these variables. Hence,
the finite temperature gap equation can be expressed as 
\begin{equation}
m(\omega _{n^{\prime }},p)=(\frac{g^{2}}{4\pi }+\frac{\lambda _{y}^{2}}{4\pi 
})\frac{T}{\pi }\sum_{n=-\infty }^{\infty }\int\limits_{-\infty }^{\infty }%
\frac{dk\text{ }m(\omega _{n},k)}{\omega _{n}^{2}+k^{2}+m^{2}(\omega _{n},k)}%
\int\limits_{0}^{\infty }\frac{dk\exp (-\frac{x}{2\left| gH\right| })}{%
(\omega _{n}-\omega _{n^{\prime }})^{2}+(k-p)^{2}+x}  \label{e12}
\end{equation}
with $\omega _{n}=(2n+1)\pi T.$ Eq. (\ref{e12}) differs from the
corresponding equation for QED (Eq.(10) of reference \cite{shoko}) only in
the factor multiplying the integrals, which in the present case contains the
contribution $\frac{\lambda _{y}^{2}}{4\pi }$ due to the Yukawa interaction.

It is natural to expect that at some critical temperature the thermal
effects evaporate the $\overline{\psi }\psi $ fermion condensate responsible
for the nonzero dynamical mass. For the dynamical mass generation to be of
any significance at the electroweak scale, it is needed that the critical
temperature at which the mass becomes zero results of the order of the
electroweak critical temperature $\sim 10^{2}Gev.$ Of course, we must
remember that there is another parameter in the problem, the magnetic field.
Different magnetic field strengths will yield different values of the
critical temperature. Therefore, our goal is to determine the magnetic field
required to have a critical temperature comparable to the electroweak one.
Hence, we need to find the critical temperature as a function of the
couplings and the magnetic field, that is, we need to solve first the gap
equation (\ref{e12}), and then take the limit $m\rightarrow 0$. Since our
calculations are very similar to those of ref.\cite{shoko}, we remit the
interested lector to that paper, and here we just give the final result,
which in our case contains the modification due to the extra coupling $%
\lambda _{y}.$ Thus, the critical temperature estimate is 
\begin{equation}
T_{c}\approx \sqrt{\left| gH\right| }\exp \left[ -\sqrt{\frac{\pi }{\frac{%
g^{2}}{4\pi }+\frac{\lambda _{y}^{2}}{4\pi }}}\right] \simeq m(T=0)
\label{n29}
\end{equation}
which results comparable to the dynamical mass at zero temperature. Critical
temperature estimates of the order of the corresponding zero temperature
dynamical masses have been also found in QED and NJL, in (2+1) and (3+1)
dimensions\cite{mirans}\cite{shoko}\cite{ebert}.

We can now estimate the strength of the magnetic field required to have $%
T_{c}\sim 10^{2}Gev.$ For Yukawa coupling $\lambda _{y}\simeq 0.7$ and gauge
coupling $g=\frac{2}{3}e$ we have 
\begin{equation}
T_{c}\simeq 10^{2}Gev\simeq 1.8\times 10^{-4}\sqrt{\left| gH\right| }\simeq
1.2\times 10^{-12}Gev\sqrt{\frac{H(G)}{10^{4}G}}  \label{n30}
\end{equation}
thus the critical field is $H\approx 10^{32}G.$ This result represents a
decreasing of the field in 10 orders of magnitude as compared to the value
required in QED (where $\lambda _{y}\simeq 0,g=e)$ to obtain the same
critical temperature of $10^{2}Gev.$

In a note added in proof, Lee et. al.\cite{lee} concluded that the dynamical
generation of mass due to a magnetic field plays no role in the electroweak
phase transition, because the field that had to be present in the early
universe for the FDGM to be important was too large. Their claim was based
in results obtained within QED$_{4}$, where the critical field strength is $%
\sim 10^{42}G$. Here we have a critical field $\sim 10^{32}G$. Although much
smaller, it is still larger than $10^{24}G$, which is the estimate generally
predicted by primordial field generating mechanisms\cite{gen-mech}. We must
point out, however, that the present result has been found within a toy
model, so it just indicates the tendency of the theory when new interactions
are switched on. We anticipate that in the electroweak model the critical
field needed shall be much smaller than the one found here.

\section{Conclusions}

The main conclusion of this paper is that Yukawa interactions enhance the
dynamical generation of fermion bound states and masses in the presence of
external magnetic fields. It is natural to expect that other interactions
will have similar consequences. Therefore, it is worth to extend our results
to more realistic theories. In the context of the electroweak model the
enhancement of the FDGM should be even more substantial due to the
interaction richness of the theory.

The generation of a fermion condensate in the presence of a constant
magnetic field provides an example of change in the symmetry properties of
the vacuum due to external fields and may have very wide applications. The
FDGM was originally discovered as a mechanism of catalysis of chiral
symmetry breaking \cite{mirans} and as so, it has been rediscovered in many
different theories, including condensed matter phenomena. We believe that in
gauge theories with a richer set of interactions and symmetries the FDGM may
be connected to the breaking of a gauge symmetry. In the context of the
Standard Model our conjecture poses an interesting question: could
primordial magnetic (hypermagnetic) fields induce gauge symmetry breaking in
the early universe? If the answer is positive, it may led to several
cosmological consequences. Work on this direction is in progress and will be
publish elsewhere.

\bigskip

\begin{description}
\item  
\begin{center}
{\bf ACKNOWLEDGMENTS}
\end{center}
\end{description}

It is a pleasure to thanks D. Caldi, V. P. Gusynin, C. N. Leung, Y. J. Ng,
and I. A. Shovkovy for very useful discussions. Our special thanks to V. A.
Miransky for enlightening discussions and for calling our attention to the
basic paper of ref.[3]. This work has been supported in part by NSF grant
PHY-9722059

\bigskip

\begin{description}
\item  
\begin{center}
{\bf APPENDIX}
\end{center}
\end{description}

Let us find the solution of $Z_{_{\Vert }}$ that satisfies the equation 
\begin{equation}
Z_{_{\Vert }}\gamma \cdot \overline{p}_{_{\Vert }}=-4\lambda _{y}^{2}(\left|
gH\right| )\int \frac{d^{4}\widehat{q}}{(2\pi )^{4}}\frac{e^{-\widehat{q}%
_{\bot }^{2}}}{\widehat{q}^{2}}\frac{(1+Z_{_{\Vert }})\gamma \cdot (%
\overline{p}_{_{\Vert }}-q_{_{_{\Vert }}})}{(1+Z_{_{\Vert }})^{2}(\overline{p%
}_{_{\Vert }}-q_{_{_{\Vert }}})^{2}+m^{2}\left( \left| \overline{p}_{_{\Vert
}}-q_{_{_{\Vert }}}\right| \right) }  \eqnum{A1}  \label{A1}
\end{equation}

Doing the variable change 
\begin{equation}
k_{_{_{\Vert }}}=\left( k_{4},k_{3}\right) =\left( \overline{p}_{4}-q_{4},%
\overline{p}_{3}-q_{3}\right)  \eqnum{A2}  \label{A2}
\end{equation}
the integral in Eq. $\left( \ref{A1}\right) $ can be written as 
\begin{eqnarray}
I &=&\int \frac{d^{4}\widehat{q}}{(2\pi )^{4}}\frac{e^{-\widehat{q}_{\bot
}^{2}}}{\widehat{q}^{2}}\frac{(1+Z_{_{\Vert }})\gamma \cdot (\overline{p}%
_{_{\Vert }}-q_{_{_{\Vert }}})}{(1+Z_{_{\Vert }})^{2}(\overline{p}_{_{\Vert
}}-q_{_{_{\Vert }}})^{2}+m^{2}\left( \left| \overline{p}_{_{\Vert
}}-q_{_{_{\Vert }}}\right| \right) }  \nonumber \\
&=&\int \frac{d^{2}\widehat{q}_{\bot }d^{2}\widehat{k}_{\Vert }}{(2\pi )^{4}}%
\frac{e^{-\widehat{q}_{\bot }^{2}}}{\widehat{q}_{\bot }^{2}+\frac{1}{2\left|
gH\right| }\left( \overline{p}_{\Vert }-k_{\Vert }\right) ^{2}}\frac{%
(1+Z_{_{\Vert }})\gamma \cdot k_{_{\Vert }}}{(1+Z_{_{\Vert }})^{2}(\overline{%
k}_{_{\Vert }})^{2}+m^{2}\left( \left| k_{_{_{\Vert }}}\right| \right) } 
\eqnum{A3}  \label{A3}
\end{eqnarray}

We can change $k_{\Vert }$ now to polar coordinates $\left( \kappa _{_{\Vert
}},\theta _{k}\right) $, and use Feynman integral formula 
\begin{equation}
\frac{1}{AB}=\int\limits_{0}^{1}dy\frac{1}{\left[ yA+\left( 1-y\right)
B\right] ^{2}}  \eqnum{A4}  \label{a4}
\end{equation}
to write the integral $\left( \ref{A3}\right) $ as 
\begin{eqnarray}
I &=&\int \frac{d^{2}\widehat{q}_{\bot }}{(2\pi )^{4}}\int\limits_{0}^{%
\infty }d\kappa _{_{\Vert }}\int\limits_{0}^{1}dy\int\limits_{0}^{2\pi
}d\theta _{k}\widehat{\kappa }_{_{\Vert }}e^{-\widehat{q}_{\bot }^{2}}\left(
1+Z_{\Vert }\right)  \nonumber \\
&&\times \frac{\gamma _{_{0}}\kappa _{_{\Vert }}\cos \theta _{k}+\gamma
_{_{3}}\kappa _{_{\Vert }}\sin \theta _{k}}{\left\{ y\widehat{q}_{\bot }^{2}+%
\frac{y}{2\left| gH\right| }\left( {\rm p}_{\Vert }^{2}+\kappa _{\Vert
}^{2}\right) -\frac{2{\rm p}_{\Vert }\kappa _{_{\Vert }}}{2\left| gH\right| }%
\cos \left( \theta _{k}-\theta _{p}\right) +\left( 1-y\right) \left[
(1+Z_{_{\Vert }})^{2}(\kappa _{_{\Vert }})^{2}+m^{2}\right] \right\} ^{2}} 
\eqnum{A5}  \label{a5}
\end{eqnarray}
where we have written the external parallel momentum in polar coordinates 
\begin{eqnarray}
\overline{p}_{4} &=&p_{4}={\rm p}_{\Vert }\cos \theta _{p}  \nonumber \\
\overline{p}_{3} &=&p_{3}={\rm p}_{\Vert }\sin \theta _{p}  \eqnum{A6}
\label{a6}
\end{eqnarray}

Using the following formulas, which are valid for $a^{2}\neq b^{2},$%
\begin{equation}
\int d\theta \frac{\cos \theta }{\left( a+b\cos \theta \right) ^{2}}=\frac{%
a\sin \theta }{\left( a^{2}-b^{2}\right) \left( a+b\cos \theta \right) }+%
\frac{1}{a^{2}-b^{2}}\int d\theta \frac{1}{a+b\cos \theta }  \eqnum{A7}
\label{a7}
\end{equation}
\begin{equation}
\int d\theta \frac{\sin \theta }{\left( a+b\cos \theta \right) ^{2}}=\frac{1%
}{\left( a^{2}-b^{2}\right) }-\frac{a}{b\left( a^{2}-b^{2}\right) }\ln
\left( a+b\cos \theta \right)  \eqnum{A8}  \label{a8}
\end{equation}
\begin{equation}
\int d\theta \frac{1}{a+b\cos \theta }=\left\{ 
\begin{array}{c}
\frac{2}{\sqrt{a^{2}-b^{2}}}arctg\frac{\left( a-b\right) tg\frac{\theta }{2}%
}{\sqrt{a^{2}-b^{2}}},\qquad if\quad a^{2}>b^{2} \\ 
\frac{1}{\sqrt{b^{2}-a^{2}}}\ln \frac{\left( a-b\right) tg\frac{\theta }{2}-%
\sqrt{b^{2}-a^{2}}}{\left( a-b\right) tg\frac{\theta }{2}+\sqrt{b^{2}-a^{2}}}%
,\qquad if\quad a^{2}<b^{2}
\end{array}
\right\}  \eqnum{A9}  \label{a9}
\end{equation}

and the definitions

\begin{equation}
a=y\widehat{q}_{\bot }^{2}+\frac{y}{2\left| gH\right| }\left( {\rm p}_{\Vert
}^{2}+\kappa _{\Vert }^{2}\right) +\left( 1-y\right) \left[ (1+Z_{_{\Vert
}})^{2}(\kappa _{_{\Vert }})^{2}+m^{2}\right]  \eqnum{A10}  \label{a10}
\end{equation}

\begin{equation}
b=-\frac{2{\rm p}_{\Vert }\kappa _{_{\Vert }}}{2\left| gH\right| }\cos
\left( \theta _{k}-\theta _{p}\right)   \eqnum{A11}  \label{a11}
\end{equation}
one can straightforwardly show that all the angle integrals are zero.
Therefore, Eq. $\left( \ref{A1}\right) $ reduces to 
\begin{equation}
Z_{_{\Vert }}\gamma \cdot \overline{p}_{_{\Vert }}=0  \eqnum{A12}
\label{a12}
\end{equation}
leading to the solution $Z_{_{\Vert }}=0.$


\begin{references}
\bibitem{mirans}  V. P. Gusynin, V. A. Miransky, and I. A. Shovkovy, Phys.
Rev. Lett., {\bf 73}, 3499 (1994); Phys.Lett. B {\bf 349}, 477 (1995); Phys.
Rev. D {\bf 52}, 4718 (1995); ibid 4747; Nucl. Phys. B {\bf 462},249 (1996).

\bibitem{ackley}  C. N. Leung, Y. J. Ng, and A. W. Ackley, Phys. Rev. D {\bf %
54}, 4181 (1996).

\bibitem{lee}  D.-S Lee, C. N. Leung, and Y. J. Ng, Phys. Rev. D {\bf 55},
6504 (1997).

\bibitem{shoko}  V. P. Gusynin and I. A. Shovkovy, Phys. Rev. D {\bf 56},
5251 (1997).

\bibitem{hongkim}  D. K. Hong, Y. Kim, and S-J. Sin, Phys. Rev. D {\bf 54},
7879 (1996).

\bibitem{hong}  D. K. Hong, Phys.Rev. D {\bf 57,} 3759 (1998).

\bibitem{ebert}  D. Ebert and V. Ch. Zhukovsky, Mod. Phys. Lett. A {\bf 12},
2567 (1997).

\bibitem{grupo}  V. Elias, D. G. C. McKeon,V. A. Miransky, and I. A.
Shovkovy, Phys. Rev D {\bf 54}, 7884 (1996); I. A. Shovkovy and V. M.
Turkovsky, Phys. Lett. B {\bf 367}, 213 (1996); D. M. Gitman, S. D.
Odintsov, and Yu. I. Shil'nov, Phys. Rev D {\bf 54}, 2968 (1996); A. V.
Shpagin, hep-ph/9611412; I. A. Shushpanov and A. V. Smilga, Phys. Lett. B 
{\bf 402}, 351 (1997); E. Elizalde, Yu. I. Shil'nov, and V. V. Chitov,
Class. Quant. Grav. {\bf 15}, 735 (1998); V. P. Gusynin, D. K. Hong, and I.
A. Shovkovy, Phys. Rev D {\bf 57}, 5230 (1998); V. A. Miransky,
hep-th/9805159, V. P. Gusynin and A. V. Smilga, hep-ph/9807486.

\bibitem{vv1}  E. J. Ferrer and V. de la Incera, hep-th/9805103, to appear
in Proc. of SILAFAE'98, April 8-11, San Juan, Puerto Rico.

\bibitem{note}  The FDGM may be important in the vicinity of magnetars.
Magnetars are a rare type of star whose existence has been recently
confirmed by astronomical observations of the largest batch of energy (in
the form of gamma light) to arrive at Earth from a star beyond our solar
system (see Kouveliotou, C, et al., International Astronomical Union
Circular No.7001 (Aug. 28) 1998). They are supposed to be responsible for
the largest magnetic field ever observed in our present universe, $\sim
10^{15}$ $G$. Such a field is about 100 times larger than ordinary neutron
star fields.

\bibitem{mavro}  K. Farakos and N. E. Mavromatos, cond-mat/9710288; K.
Farakos, G. Koutsoumbas, and N. E. Mavromatos, Phys.Lett. B {\bf 431,} 147
(1998); cond-mat/9805402.

\bibitem{semenof}  G. W. Semenoff, I. A. Shovkovy, and L. C. R.
Wijewardhana, hep-ph/9803371.

\bibitem{gen-mech}  T. Vachaspati, Phys.Lett. B {\bf 265},258 (1991);K.
Enquist and P. Olesen, ibid.{\bf 319}, 178 (1993); B. Cheng and Olinto,
Phys. Rev. D {\bf 50,} 2421 (1994); G. Baym, D. B\"{o}deker, and McLerran,
ibid.{\bf 53},662 (1996); P. Olesen. hep-ph/9708320, A. Brandenburg, K.
Enquist, and P. Olesen, Phys. Rev. D {\bf 54}, 1291 (1996); Phys.Lett. B 
{\bf 392}, 395 (1997); K. Enquist,astro-ph/9707300, G. Sigl, A. Olinto, and
K Jedamzik, Phys. Rev. D {\bf 55}, 4582 (1997); M. Joyce and M.
Shaposhnikov, Phys. Rev. Lett. {\bf 79},1193 (1997); M. Giovannini and M.
Shaposhnikov,Phys. Rev. D {\bf 57}, 2186 (1998); O. T\"{o}rnkvist,
Phys.Rev.D {\bf 58}:043501, (1998); hep-ph/9801286; M. Hindmarsh and A.
Everett, astro-ph/9708004; D. Grasso and A. Riotto, Phys. Lett. B {\bf 418},
258 (1998); J. Ahonen and K.Enquist, Phys. Rev. D 57. 664 (1998).

\bibitem{olesen}  J. Ambjorn and P. Olesen, hep-ph/9304220.

\bibitem{vv}  E. J. Ferrer and V. de la Incera, Phys. Rev D {\bf 58}, 065008
(1998).

\bibitem{schwinger}  J. Schwinger, Phys. Rev. {\bf 82}, 664 (1951)

\bibitem{ritus}  V. I. Ritus in Issues in Intense-Field Quantum
Electrodynamics, ed. V. L. Ginzburg (Nova Science, Commack, 1987).

\bibitem{handbook}  {\it Handbook of Mathematical Functions}, eds. M.
Abramowitz and I. A. Stegun (Dover, New York, 196).

\bibitem{note2}  A straighforward calculation can convince us that the
fermion condensate in the model considered in this work has the same
expression as the fermion condensate found in ref. [3], that is, $%
\left\langle \overline{\psi }\psi \right\rangle =iTr\left\{ \overline{G}%
(x,x)\right\} \simeq -\frac{\left| gH\right| }{2\pi ^{2}}m\ln \left( \frac{%
\left| gH\right| }{m^{2}}\right) $. The difference now is that the present
mass is given by Eq. $\left( \ref{n28}\right) $, so the condensate is
enhanced by the Yukawa coupling.
\end{references}
\end{document}